\documentclass[twocolumn,conference]{IEEEtran}
\usepackage{amsmath,amscd,graphicx,latexsym,multicol,cite,ulem,epsf}
\usepackage{verbatim}
\usepackage{graphics}
\usepackage{amssymb}

\begin{document}
\title{Hierarchical Resource Allocation in Femtocell Networks using Graph Algorithms}
\author{Sanam Sadr and Raviraj Adve\\
University of Toronto \\
Dept. of Elec. and Comp. Eng.,  \\ 10 King's College Road, Toronto,
Ontario, Canada M5S 3G4\\
\{ssadr, rsadve\}@comm.utoronto.ca }
\date{}
\maketitle
\begin{abstract}
This paper presents a hierarchical approach to resource allocation in
open-access femtocell networks. The major challenge in femtocell
networks is interference management which in our system, based on the
Long Term Evolution (LTE) standard, translates to which user should be
allocated which physical resource block (or fraction thereof) from
which femtocell access point (FAP). The globally optimal solution
requires integer programming and is mathematically intractable. We
propose a hierarchical three-stage solution: first, the load of each
FAP is estimated considering the number of users connected to the FAP,
their average channel gain and required data rates. Second, based on
each FAP's load, the physical resource blocks (PRBs) are allocated to FAPs in a manner that minimizes the interference by coloring the modified interference graph. Finally, the resource allocation is performed at each FAP considering
users' instantaneous channel gain. The two major advantages of this
suboptimal approach are the significantly reduced computation
complexity and the fact that the proposed algorithm only uses
information that is already likely to be available at the nodes
executing the relevant optimization step. The performance of the
proposed solution is evaluated in networks based on the LTE standard.
\end{abstract}

\section{Introduction}
Each new generation of wireless communication systems promises higher
quality of service to a larger number of users. This requires new
systems to find greater efficiencies in use of the scarce resources,
especially, the radio spectrum. In the cellular context, an effective
approach is to reduce the distance between the transmitter and the
receiver, in turn reducing the transmit power and increasing the
frequency reuse factor. Coupled with the observed increase in the
indoor data activity~\cite{ABI:07}, femtocells have been proposed as
user-deployed small base stations to improve indoor coverage. The idea
is that the user equipment (UE) would hand off from the cellular base
station to the femtocell access point (FAP) installed at home or an
office when operating indoors or when it is close enough to the FAP.
The potential increase in the system capacity, however, encourages
investigations into deploying femtocells to also service any set of
opportunistic users which "see" the FAP (generally called open
access)~\cite{andrews:08}.

For FAPs to service outdoor users, they must transmit adequate power to
provide coverage to a larger area. This, in turn, results in higher
interference if neighbouring FAPs share a channel. In a
multi-transmitter environment, such interference would dominate
performance. Hence, several methods have been proposed in literature to
reduce interference. They mainly fall into two categories: power
auto-configuration e.g.,~\cite{claussen:08} and cognitive and dynamic
spectrum allocation~\cite{perez:09, sousa:09, zhang:10} (amongst many
references). However, these works either require a lot of information
to execute the relevant algorithm or require completely decentralized
decision making that leads to interference or long convergence times to
a solution.

In this paper, we focus on interference avoidance while allocating
resources to meet users' minimum rate demands. Our analysis and design
is in the context of femtocell-assisted networks based on the Long Term
Evolution (LTE) standard. Here, resource allocation is defined in terms
of connecting users to FAPs, the allocation of physical resource blocks
(PRBs) to FAPs based on the users it supports and then to users.
Clearly, solving this resource allocation problem to obtain the
globally optimal solution requires integer programming and is,
essentially, impossible in practice. In response, we present a
partially decentralized and hierarchical resource allocation process.
Each step in the process uses only local information, i.e., information
is likely to be easily obtained at the node executing the step. Since
LTE allows for reallocation of PRBs every millisecond~\cite{rohde:07},
we allow for fractional allocation of PRBs to users; the fraction
indicates the fraction of time the user is allocated the PRB.

The overall algorithm comprises three steps: (i) deciding the PRB
requirements at FAPs based on users' data requirements and
\emph{average} channel gains; (ii) allocating specific PRBs to FAPs
based on coloring an interference graph; (iii) FAPs determining max-min
user allocations based on their required data rates. This work is
presented in the downlink, however, similar work can be done in the
uplink with different power constraints introduced to the problem. The
key contribution of this paper is, therefore, a partially distributed
algorithm that provides effective resource allocation and can be scaled
to a large number of FAPs and users.

The rest of the paper is organized as follows: In
Section~\ref{system}, we describe the system model and formulate the
global optimization problem. The proposed sub-optimal solution is
presented in Section~\ref{proposed}. The simulation results are
performed in LTE standard presented in Section~\ref{results}. The paper
concludes with a discussion of the results in Section~\ref{conclusion}.

\section{System Model and the Objective} \label{system}
\begin{figure}
\center
\includegraphics [scale = 0.24]{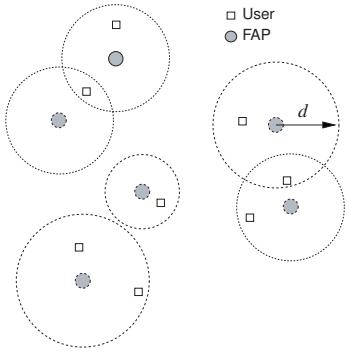}
\caption{Random distribution of FAPs in a femtocell Network.}
\label{fig:femtonet}
\end{figure}

Figure~\ref{fig:femtonet} illustrates the femtocell network under
consideration. We assume a single circular cell and independent uniform
distribution for the location of all users and FAPs. A user connects to
a single FAP only. We do not consider the basestation (BS) since it is
generally provided an orthogonal frequency/PRB allocation; our model
considers only users using FAPs\footnote{If the BS allocation is not
orthogonal, it can be easily incorporated as another transmitting node
in the network with its own power budget.}. Each femtocell access point
is at the centre of its coverage area. In the figure, $d$ is the radius
of the coverage area which depends on the environment and the FAP
transmit power. Due to random geographic distribution of femtocells,
some of them might interfere. In other words, some users might be in a
location covered by more than one FAP.

The system comprises $K$ users, and $L$ FAPs are assumed according to a
chosen user and FAP density. The physical channel between FAPs and
users is modelled as frequency selective Rayleigh fading with average
power determined by distance attenuation and large scale fading
statistics. The total bandwidth is $B$ subdivided into $N$ sub-channels
or PRBs. User $k$ has a minimum data requirement of $R_k$ bits per
second (bps). To achieve fairness, we aim to maximize the minimum
proportional rate, i.e., max-min of the achieved rate over all the
users normalized by their required data rates. For convenience, the
rate achieved on a specific channel is assumed to be given by the
Shannon capacity, though more practical modulations can be easily
accounted for using the corresponding gap to
capacity~\cite{goldsmith:97}.

Under this setting, the general form of the optimization problem is:
\begin{eqnarray}
\label{prob1}
\max_{p_{k,n}^{(l)}, S_l} \min_{k} & & \hspace*{-0.2in} \frac{1}{R_k}\left[\sum_{n = 1}^{N}\frac{B}{N}
    \log_{2}\left ( 1 +  \frac{p_{k,n}^{(l)} h_{k,n}^{(l)}}
    {\sum_{i \neq l}p_{k,n}^{(i)} h_{k,n}^{(i)} + \sigma^{2}} \right )\right] \nonumber \\
\mbox{subject to} & & \hspace*{-0.2in} \sum_{n =1}^{N}\sum_{k \in S_{l}}p_{k,n}^{(l)} \leq P_{max},
                                                    \; \; \;  l=1,2, ..., L \nonumber\\
& & \hspace*{-0.2in} p_{k,n}^{(l)} \geq 0, \; \; \forall k,n \\
& & \hspace*{-0.2in} \bigcup_{l =1}^{L} S_{i} = K, \; \; \; S_{i} \cap S_{j} =
                                                    \emptyset \; \; i \neq j \nonumber
\end{eqnarray}
where $S_{l}$ is the set of users connected to and being serviced by
FAP $l$. $p_{k,n}^{(l)}$ and $h_{k,n}^{(l)}$ are, respectively, the
assigned power and channel power gain from FAP $l$ to user $k$ on
subchannel $n$. $B/N$ is the bandwidth and $\sigma^{2}$ the noise power
on each subchannel\footnote{The terms ``PRB" and ``subchannels" are
used interchangeably.}. The first constraint is on the total transmit
power of each FAP, while the second ensures non-negative powers.  The
third constraint ensures all users are connected to an FAP. Finally,
the sets, $\left\{S_{l}\right\}_{l = 1}^L$ are disjoint since each user
is serviced by one and only one FAP at a time.

To solve this problem, we need to find the optimal $S_{l}$ and
$p_{k,n}^{l}$ leading to the max-min normalized rates. There are three
issues that make finding the solution to this optimization problem
impossible in practice: one, the assignment of users to FAPs makes this
an intractable integer programming problem; two, the problem is
non-convex and hence hard to solve; three, solving this optimization
problem requires knowledge of all subchannels for all the users to all
the FAPs. Getting this information to a central server would involve a
huge overhead. Essentially, a resource allocation scheme based on
perfect knowledge of channels is infeasible in a femtocell network.

\section{Proposed Hierarchical Solution}\label{proposed}

This section presents the proposed algorithm that makes a series of
reasonable, if sub-optimal, simplifications. First, we set the user-FAP
connections by forcing each user to request service from the one FAP
which offers the highest long term \emph{average} received power
(based, e.g., on a pilot and large-scale fading). This addresses the
first issue raised above and is a significant step in making useful
solutions feasible. We then decompose the problem of resource
allocation into three parts: (i) load estimation, (ii) PRB allocation
among FAPs and (iii) resource allocation among users at each FAP. Two
of these phases are done locally at the FAP with small number of users
and/or subchannels. This hierarchal decomposition enables us to take
advantage of locally optimal algorithms.

\subsection{Phase 1: Load Estimation}
The FAP is assumed aware of the desired rate and \emph{average channel
power} for all users that it serves. Based on this information, the FAP
\emph{roughly estimates} its load in the form of total number of
subchannels required by its users. At this phase, the objective at each
FAP is to minimize the total number of required subchannels. This
problem can be formulated as a convex optimization problem at each FAP
as:
\begin{eqnarray}
\label{prob2}
\min_{w_{k}, P_k}& & \sum_{k \in S_{l}} w_{k} \nonumber \\
\mbox{subject to} & &  w_{k}\frac{B}{N}\log_{2}\left ( 1 +
    \frac{P_{k} H_{k}}{w_{k}\sigma^{2}} \right) \geq R_{k}, \; \; \forall k \in S_{l}  \\
&& \sum_{k \in S_{l}}P_{k} \leq P_{max}, \nonumber \\
&& P_{k} \geq 0, w_{k} \geq 0, \;\; \forall k \in S_{l} \nonumber
\end{eqnarray}
where $w_k$ is the number (can be a fraction) of subchannels FAP $l$
assigns to user $k \in S_l$. $P_{k}$ and $H_{k}$ are the total power
allocated to user $k$ and the average channel gain seen by user $k$
respectively. FAP $l$ sets $N_{l} = \sum_{k \in S_{l}}w_{k}$ as the
total number of subchannels it requires.

\subsection{Phase 2: Channel Allocation Among FAPs Using Graphs}
\begin{figure}
\center
\includegraphics [scale = 0.23]{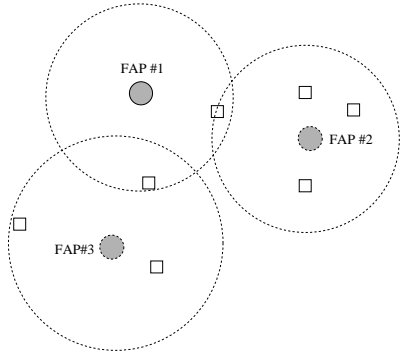}
\caption{FAP network with 3 FAPs. $N_1 = 1, N_2 = N_3 = 3$.}
\label{fig:interference}
\end{figure}

\begin{figure}
\center
\includegraphics [scale = 0.25]{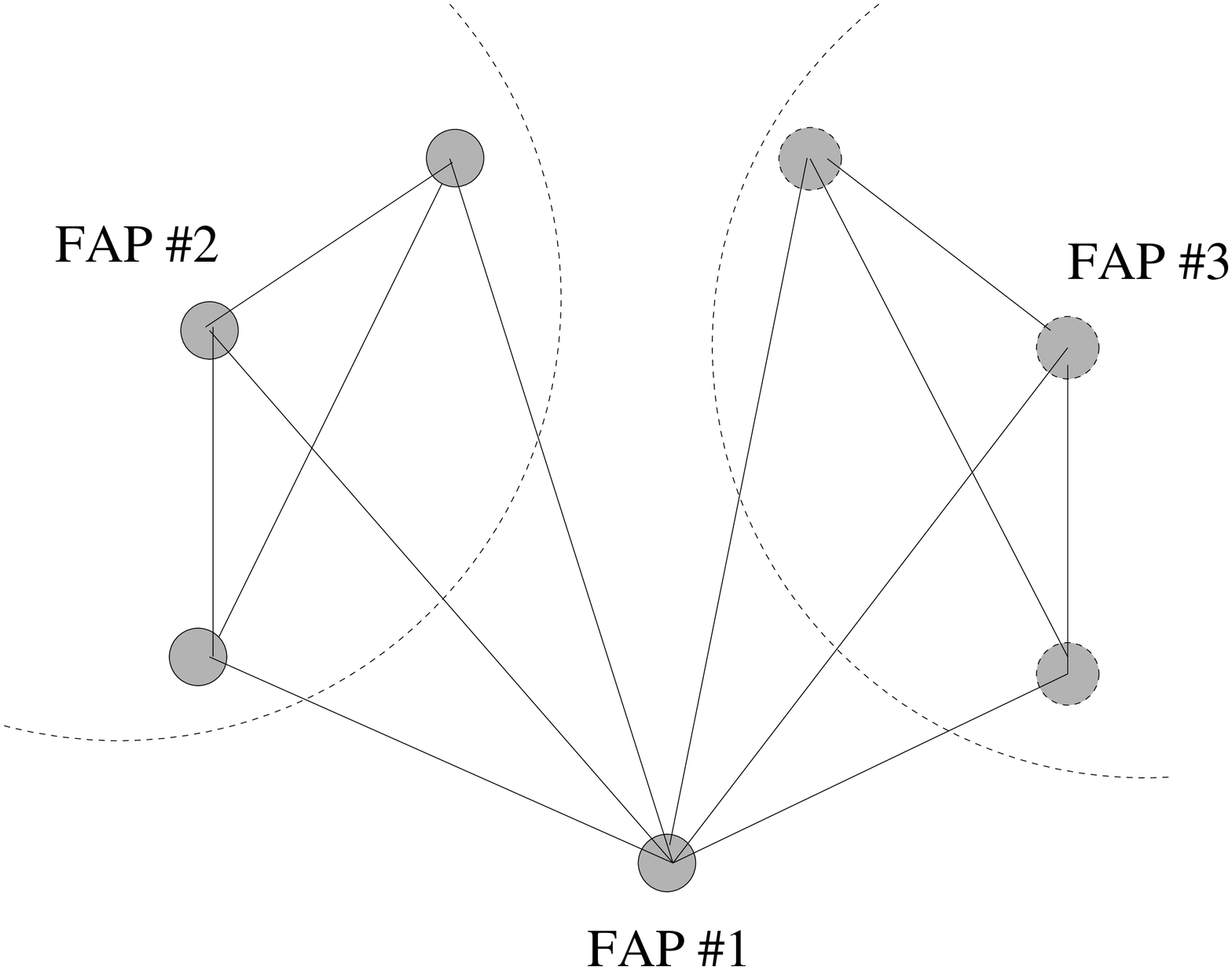}
\caption{Interference graph corresponding to Figure~\ref{fig:interference}\label{fig:inter_graph}.}
\end{figure}

\begin{figure}[h]
\center
\includegraphics [scale = 0.4]{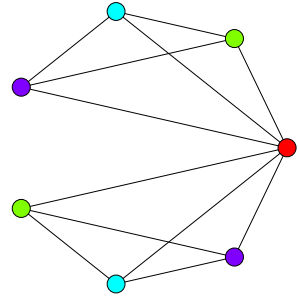}
\caption{Interference graph coloring corresponding to Figure~\ref{fig:interference}.
Minimum number of colors is 4, with both optimal and suboptimal coloring algorithms. }
\label{fig:coloring}
\end{figure}

In this phase, the objective at the server is to assign subchannels to
FAPs \emph{proportional to their estimated load} while minimizing the
interference among them. The FAPs report their requested loads,
$\left\{N_l\right\}_{l=1}^L$ to a central server. Given $N$ PRBs, if $N
\geq \sum_{l=1}^L N_l$, each FAP is easily satisfied; however, this is
unlikely. The server is assumed to know the interference graph based on
\emph{large-scale} statistics, i.e., it knows which FAPs potentially
interfere with each other. The problem of channel assignment among FAPs
can then be addressed by coloring the interference graph; here, each
color corresponds to one subchannel. To take into account the load of
FAPs, we modify the interference graph as follows: each FAP $l$ is
represented by $N_{l}$ nodes (forming a complete subgraph). An edge
connects two nodes if FAPs potentially interfere. The problem of
channel assignment then becomes a graph coloring problem where two
interfering nodes should not be assigned the same color. An example of
a three FAP network is illustrated in Fig.~\ref{fig:interference}. FAP
\#1 potentially interferes with FAP \#2 and FAP \#3. The corresponding
interference graph and the coloring are illustrated in
Figs.~\ref{fig:inter_graph} and~\ref{fig:coloring}.

For arbitrary graphs, graph coloring is an NP hard problem. We adopt
the heuristic algorithm proposed by Br\'{e}laz~\cite{Brelaz:79}; at
every iteration, the vertex which is adjacent to the greatest number of
differentely-colored neighbours is colored with a new color, if necessary
(until colors are exhausted). Note that by creating $N_l$ nodes for FAP
$l$, this process achieves an approximate proportional fairness in the
case the server cannot meet the demand of $\sum_l N_l$
interference-free subchannels.

It is worth noting that optimal coloring of graphs is possible with low
complexity if the interference graph is sparse such that each node is
connected to at most $N$ nodes where $N$ is the total number of
channels. Such graphs can be optimally colored with a modified Breadth
First Search (BFS) algorithm with complexity of $\mathcal{O}(|V| +
|E|)$ where $|E| = \alpha |V| = \mathcal{O}(|V|)$ where $|E|$ and $|V|$
are the cardinality of edges and vertices respectively and $\alpha$ is
a scalar. Furthermore, due to the small coverage area of each
femtocell, we assume FAPs are either interfering or not. However, by
changing the interference graph into a weighted graph, the problem of
interference reduction can also be formulated as max $K$-Cut problem
where $N$ is the number of available subchannels. The goal then is to
find $N$ groups of nodes where each group has the lowest edge weight
among them and can be assigned the same color.

\subsection{Phase 3: Resource Allocation Among Users}
At the end of Phase 2, the FAPs have been assigned their subchannels.
Each FAP can now allocate these resources amongst the users it serves.
At each FAP, the problem is now reduced to maximizing the minimum rate
for each user subject to their requested data rate. The FAP
is assumed to know the channel state information for the users it
serves. Let $\bar{N}_l$ be the number of subchannels assigned to FAP
$l$  (not necessarily equal to $N_l$, the number of subchannels
requested). At this FAP, the resource allocation problem is formulated
as:
\begin{eqnarray}
\label{prob3}
\max_{p_{k,n}, c_{k,n}} \min_{k} & & \frac{1}{R_k}\frac{B}{N}\sum_{n = 1}^{\bar{N}_{l}}c_{k,n}
    \log_{2}\left ( 1 +  \frac{p_{k,n} h_{k,n}}{\sigma^{2}} \right )
    \nonumber \\
\mbox{subject to} & & \sum_{n =1}^{\bar{N}_{l}}\sum_{k \in S_{l}}p_{k,n} \leq P_{max}, \hspace*{0.2in}
p_{k,n} \geq 0, \; \; \forall k,n \nonumber \\
& & \sum_{k \in S{l}} c_{k,n} = 1, \; c_{k,n} \geq 0 \; \; \forall n
\end{eqnarray}
where $c_{k,n}$ is the fraction of subchannel $n$ (using time-division
as allowed by LTE) allocated to user $k$.

This is a standard convex optimization problem easily solved for the
time fraction and assigned powers. An even simpler alternative is to
divide power equally amongst the $\bar{N}_l$ subchannels, leading to a
linear program.

To summarize the steps of the algorithm, the FAP reports its desired
subchannel allocation to a central server based on long-term
statistics. The server then uses the interference graph to assign
subchannels in an interference free manner. Finally, the FAPs use their
allocation optimally (locally) to provide resources to users. At each
stage, the problems being solved use only local knowledge.

\section{Simulation Results}\label{results}
In this section, we evaluate the performance of the proposed scheme
with simulations. The simulation setup is the LTE standard closely
following~\cite{forum}. The path loss between the FAP and the user
accounts for indoor and outdoor propagation:
\[
{\rm PL} = 38.46 + 20\log_{10}(d_{in}) + 37.6\log_{10} (d) + L + L_s \]
where $d_{in}$ is the distance between the FAP and the external wall or
window and has a uniform distribution between 1m and 5m; $L$ is the
penetration loss and is set to 10dB and 3dB (with equal probability)
for an external wall and windows respectively; $L_s$ accounts for
shadowing and is modeled as a log-normal random variable with standard
deviation of 10dB. Finally, assuming Rayleigh fading, the instantaneous
power of the received signal is modeled by an exponential random
variable with the mean equal to the average receive
power~\cite{hansen:77}. The receiver noise power spectral density is
set to -174dBm/Hz with an additional noise figure at the receiver.

The downlink transmission scheme for an LTE system is based on OFDMA
where the available spectrum is divided into multiple subcarriers each
of bandwidth 15kHz. Resources are allocated to users in PRBs of 12
subcarriers; hence the bandwidth of each subchannel is 180kHz and is
used as the signal bandwidth BW in calculating the noise power. Each
PRB is allocated to a user at a time for a \emph{subframe} duration of
1ms. Here we consider the maximum (20MHz) bandwidth corresponding to
100 PRBs of which $N$ PRBs are allocated to the femtocells, i.e.,
$\sum_l \bar{N}_l \leq N$. A key assumption in the simulations is that
since a PRB comprises 12 subcarriers, each PRB experiences an
independent fade, i.e., we assume that the multipath environment is
such that the fading is effectively flat for the 12 subcarriers in a
PRB, but rich enough to yield an independent fade on each PRB.

Table~\ref{tab:SimulationParameters} lists the parameters used in the
simulations, unless otherwise specified.
\begin{table}[htp]
\center
\caption{Simulation Parameters\label{tab:SimulationParameters}}
\begin{tabular}{| c | c |}
\hline
Parameters & Value \\
\hline \hline
Carrier frequency & 2 GHz  \\
Channel bandwidth & 20 MHz \\
Carrier spacing & 15 kHz \\
Resource Block & 180 kHz \\
Total Number of PRBs & 100 \\
N & 50\\
\hline
Transmit Power & 20dBm \\
Antenna gain & 0dB \\
Antenna & 1 x 1\\
Configuration & \\
\hline
Noise Figure in UE & 10dB\\
\hline
Minimum distance & 1 meter from FAP \\
 Penetration Loss & 10dB/3dB \\
(wall/window) & \\
\hline
$d$ & 15m \\ \hline
\end{tabular}
\end{table}

The interference graph is generated based on the distance between each
pair of FAPs. If the distance $D \leq 2d$, where $d$ is the radius of
the coverage circle of one FAP, the pair are connected in the graph and
are assumed to interfere with each other. In practice, whether two FAPs
interfere can be more accurately estimated by each FAP and reported to the
central server.

\begin{figure}
\centering
\includegraphics [scale = 0.45]{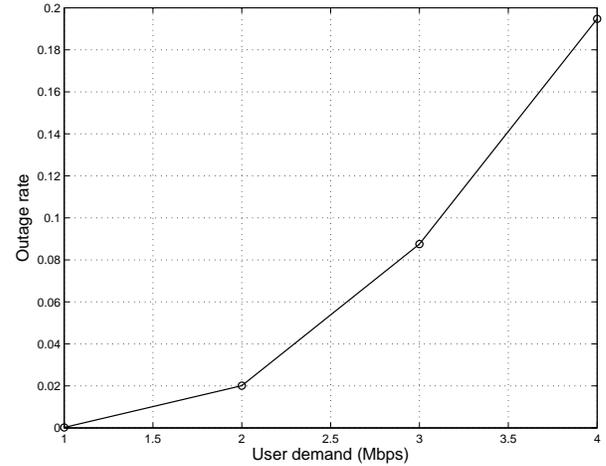}
\caption{Outage versus user rate demand in high FAP density network of
1 per circle of 5.5m radius. Cell radius = 100m.}
\label{fig:outage_vs_rate_highd}
\end{figure}

\begin{figure}
\centering
\includegraphics [scale = 0.45]{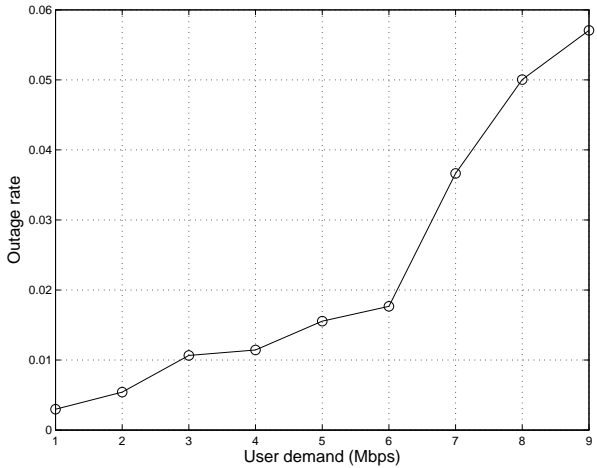}
\caption{Outage versus required data rate in low FAP density network of 1 per circle
of 18m radius. Cell radius = 100m.}
\label{fig:outage_vs_rate_lowd}
\end{figure}

\noindent \underline{\emph{Outage versus demand}}:
Figure~\ref{fig:outage_vs_rate_highd} plots the outage rate versus the
user required data rate (demand). Here outage is defined as the fraction of users who do not
receive 80\% of their requested rate. The cell radius is set to 100m
and the FAP density $\lambda$ is one per 100 $m^{2}$ equivalent to 1
per radius of 5.5m. The same metric in a low FAP density network is
shown in Figure~\ref{fig:outage_vs_rate_lowd} where $\lambda$ is 1 per
1000 $m^{2}$ equivalent to 1 FAP per radius of 18m. In both cases, user
density is chosen as $4 \lambda$ equivalent to 4 users per FAP \emph{on
average}. All the users have equal demand. The results are averaged
over 100 different user and FAP locations and 10
channel realizations per set of locations.

Comparing the two figures, the outage with high FAP density is higher
for the same user demand. This is because, in both systems, the average
number of users per FAP is constant. However, due to higher density of
FAPs, they are more likely to interfere. Hence, the interference graph
becomes dense in turn requiring more channels to color the graph. However,
the total number of PRBs is limited ($N = 50$). As a result, each FAP
gets smaller proportion of PRBs leading to higher outage.

It is worth mentioning that the key contribution here is the ability to
wisely allocate resource allocation for a large number of users, FAPs
and PRBs. In Fig.~\ref{fig:outage_vs_rate_highd}, $K=1256$, $L=314$ and
$N = 50$. Given these large numbers, it is difficult to compare these
results against any alternative scheme.

\begin{figure}[htb]
\begin{center}
\includegraphics [scale = 0.45]{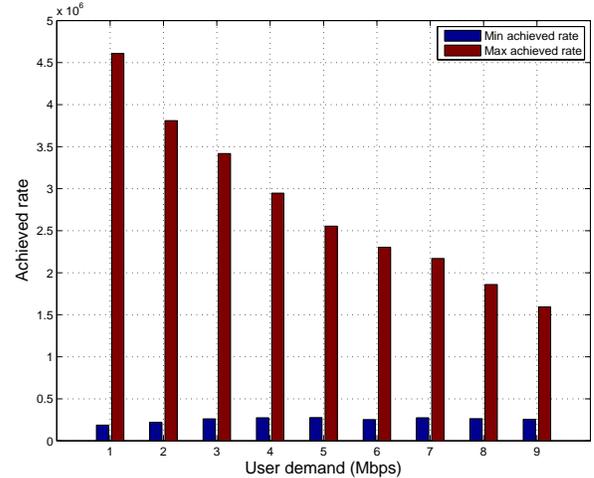}
\caption{Minimum and maximum user achieved data rate versus user demand. FAP density is
1 per circle of 18m. Cell radius = 80m.}
\label{fig:capacity_vs_rate_lowd}
\end{center}
\end{figure}

\noindent \underline{\emph{Achieved rate versus demand:}}
Figure~\ref{fig:capacity_vs_rate_lowd} plots the average minimum and
maximum (over locations and channel realizations) user data rates in a
low density network. Interestingly, the minimum achieved rate is
approximately constant with increase in the demand whereas the maximum
achieved rate decreases. This is because as the demand increases, so
does the load of each FAP and hence $N_{l}$. Since each FAP is
represented with $N_l$ virtual nodes, the resulting graph gets dense.
With the total number of PRBs fixed, each FAP receives a decreasing
fraction of its requested load. Since the local resource allocation is
in the form of max-min, the minimum data rate remains almost constant
but the maximum achieved data rate decreases. Similar results were
achieved for high density networks showing more fair distribution of
resources with increase in the all users' demand.

\section{conclusion}\label{conclusion}
In this paper, we proposed a hierarchical low complexity three-phase
resource allocation scheme in open access femtocell networks. The main
advantage of the proposed scheme is decomposing a complex non-convex
optimization problem into several smaller convex problems with smaller
sets of variables to optimize. The resulting hierarchical scheme is
effective with a large problem size. The first phase of the proposed
scheme is cell selection based on the long term averaged received power
at each user and load estimation at each FAP based on its number of
users and required service. The second phase attempts to reduce the
interference among FAPs by coloring the modified interference graph
considering FAPs load. Crucially, this step allocates resources to meet
demands at individual FAPs. Finally, a convex problem in the form of
$\max \min$ is formulated at each FAP with small set of variables to
maximize the minimum achieved data rate at each FAP.

At each step in the process, the node solving the related optimization
problem requires information that it is likely to have. The central server
only requires knowledge of the demands made by the FAP. Similarly, the
FAPs require knowledge of local channels to users they serve;
importantly, they do not require any global information. Note that the
final step of resource allocation at the FAPs would, in practice, be
solved in parallel.

\bibliographystyle{ieeetr}
\bibliography{ref2_2}
\end{document}